\documentclass[pra,aps,twocolumn,superscriptaddress,nofootinbib]{revtex4-2}
\usepackage{graphicx} 
\usepackage{amsmath}
\usepackage{xcolor}
\usepackage{amssymb}
\usepackage{physics}
\usepackage{siunitx}
\usepackage{textgreek}
\usepackage[title]{appendix}
\usepackage[uline]{hhtensor}

\renewcommand\op[1]{\mathrm{#1}}
\newcommand\Id{\mathrm{I}}
\providecommand\jbar{\bar{j}}
\providecommand\mbar{\bar{m}}
\providecommand\barlambda{\bar{\lambda}}
\providecommand\intdkmeasure{\int_{>0}^\infty {\mathrm{d}k}\,k}
\providecommand\Eq[1]{\text{Eq.~(\ref{#1})}}
\providecommand\ketbath{\ket{\Psi_\text{bath}}}
\providecommand\intdpinvmz{\int_{\mathbb{R}^3-\zerovec} \frac{\text{d}^3 \pp}{k}\text{ }}
\providecommand\zerovec{\mathbf{0}}
\providecommand\pp{\mathbf{k}}
\providecommand\phat{\mathbf{\hat{k}}}
\providecommand\ahat{\mathbf{\hat{a}}}
\providecommand\bhat{\mathbf{\hat{b}}}
\renewcommand\qq{\mathbf{q}}
\providecommand\cz{c_0}
\newcommand\exval[1]{\mathbb{E}\left\{#1\right\}}
\newcommand\ii{\mathrm{i}}
\newcommand\ketobject{\ket{\Phi_\text{thermal}}}
\providecommand\ffstar[1]{\mathrm{f}^*_{#1}(\pp)}
\providecommand\FFstar[1]{\mathrm{f}^*_{jm#1}(k)}
\providecommand\ff[1]{\mathrm{f}_{#1}(\pp)}
\providecommand\FF[1]{\mathrm{f}_{jm#1}(k)}
\providecommand\Ert{\mathbf{E}(\rr,t)}
\providecommand\rr{\mathbf{r}}
\providecommand\epsz{\varepsilon_{0}}
\providecommand\rhat{\mathbf{\hat{r}}}
\providecommand\inout{\text{in/out}}

\newcommand*\pkg[1]{\textsc{#1}}
\DeclareMathOperator\arctantwo{arctan2}

\DeclareSIUnit{\invcm}{\cm\tothe{-1}}

\begin{document}

\title{Studying thermal radiation with T-matrices}

\author{Juan Diego Mazo-V\'asquez}
\email{juan-diego.mazo-vasquez@mpl.mpg.de}
\affiliation{Institute of Theoretical Solid State Physics, Karlsruhe Institute of Technology, Kaiserstr. 12, 76131 Karlsruhe, Germany}
\affiliation{Max Planck Institute for the Science of Light, Staudtstr. 2, 91058 Erlangen, Germany}
\affiliation{Department of Physics, Friedrich-Alexander-Universität Erlangen-Nürnberg, Staudtstr. 7, 91058 Erlangen, Germany}
\author{Markus Nyman}
\affiliation{Institute of Nanotechnology, Karlsruhe Institute of Technology, Kaiserstr. 12, 76131 Karlsruhe, Germany}
\author{Marjan Krsti\'c}
\affiliation{Institute of Theoretical Solid State Physics, Karlsruhe Institute of Technology, Kaiserstr. 12, 76131 Karlsruhe, Germany}
\author{Lukas Rebholz}
\affiliation{Institute of Theoretical Solid State Physics, Karlsruhe Institute of Technology, Kaiserstr. 12, 76131 Karlsruhe, Germany}
\author{Carsten Rockstuhl}
\affiliation{Institute of Theoretical Solid State Physics, Karlsruhe Institute of Technology, Kaiserstr. 12, 76131 Karlsruhe, Germany}
\affiliation{Institute of Nanotechnology, Karlsruhe Institute of Technology, Kaiserstr. 12, 76131 Karlsruhe, Germany}
\author{Ivan Fernandez-Corbaton}
\email{ivan.fernandez-corbaton@kit.edu}
\affiliation{Institute of Nanotechnology, Karlsruhe Institute of Technology, Kaiserstr. 12, 76131 Karlsruhe, Germany}

\begin{abstract}
	We introduce a basic formalism for computing thermal radiation by combining Waterman's T-matrix method with an algebraic approach to light-matter interactions. The formalism applies to nano-particles, clusters thereof, and also molecules. In exemplary applications, we explore how a chiral structure can induce an imbalance in the circular polarization of thermal radiation. While the imbalance is rather small for a chiral molecule such as R-BINOL, a much larger imbalance is observed for an optimized silver helix of approximately \SI{200}{\nano\meter} in size. Besides the directional Kirchhoff law used in this article, the formalism is suitable for implementing more nuanced theories, and it provides a straightforward path to the computation of thermal radiation spectra of astronomical objects moving at relativistic speeds with respect to the measurement devices.  
\end{abstract}

\maketitle

\section{Introduction and summary}
Modern micro- and nanofabrication techniques extend the possibilities for achieving functional photonic systems: From bulk materials structured at the micrometer scale to molecular films and atomic monolayers. Such opportunities are exploited to control electromagnetic radiation. Thermal radiation, which is the electromagnetic radiation emitted by a body due to its temperature, is ubiquitous, and its control has particularly important applications. Examples include energy harvesting with thermophotovoltaic devices and radiative cooling \cite{Song2015,Cuevas2018,Picardi2023}. 

The design of photonic systems to control thermal radiation benefits from theoretical and numerical tools able to handle the complex possibilities enabled by micro- and nanofabrication, and exploit their potential. While the Green tensor formalism can yield semi-analytical solutions for basic systems, such as planar multilayers \cite{Joulain2005}, numerical methods are usually required as soon as more intricate geometries are involved. Several computational approaches exist to study thermal radiation, a review of which can be found in \cite[Sec.~III]{Song2015}. Among the more general methods, we can find the scattering operator approaches \cite{Krueger2012}, and boundary element methods for thermal radiation, including some with publicly available implementations \cite{Reid2009,Rodriguez2013}. 

Most computational approaches can be decomposed into two parts: A means of solving light-matter interactions, that is, scattering and absorption of electromagnetic waves by material objects, and a means of specifying the random sources or the random fields inherent to thermal radiation. A common choice for the second part is the theory of fluctuation electrodynamics started by Rytov \cite{Rytov1959}, where random sources in the objects are suitably specified. Another possibility is the common assumption, already made by Kirchhoff (see \cite[\S~2]{Baltes1976}), that the thermal emission spectrum of an object at a given temperature is independent of the environment, and can hence be computed as if the object was immersed in and in thermal equilibrium with a Planckian radiation bath at that temperature. This holds even though the object may be heated by other means, such as a warm plate or a laser beam. We will use this latter assumption in combination with Waterman's T-matrix, or transition matrix, to solve the light-matter interactions. 

The T-matrix method is a powerful and popular formalism for computing light-matter interactions. After its establishment by Waterman in 1965 \cite{Waterman1965}, the formalism has been continuously developed and improved, resulting in a large body of literature \cite{Gouesbet2019,Mishchenko2020}, and a growing number of publicly available computer codes \cite{Wriedt2008a,Hellmers2009} produced and actively maintained by the community. A distinct advantage of the T-matrix method is its general applicability. It applies to objects of arbitrary shape and material parameters, and also to molecules. The difference lies in how the T-matrices are obtained. In the first case, generic solvers of the macroscopic Maxwell equations can be used to obtain T-matrices by performing scattering calculations using the appropriate constitutive relations to describe materials \cite{Doicu1999,Mackowski2002,Fruhnert2016b,Hohenester2025}. In the case of molecules, the dynamic polarizabilities obtained through quantum-chemical calculations are mapped to T-matrices in the dipolar approximation \cite{FerCor2018}. As we will show, this allows one to compute the thermal radiation from molecules. From all the existing methods, the T-matrix approach to thermal radiation presented here is most related to the work of Kr\"uger {\em et al.} \cite{Krueger2012}, which uses the $\mathbb{T}$ operator within the Green tensor setting. We note that Waterman's T-matrix, which we use, is equal to $\mathbb{G}_0\mathbb{T}$ in \cite{Krueger2012}, namely the product of the Green tensor of the empty space $\mathbb{G}_0$, and the $\mathbb{T}$ operator. 

Here, we put forward a basic formalism for computing thermal radiation spectra by combining the T-matrix method with an algebraic approach to light-matter interactions \cite{Gross1964,Birula1996,FerCor2024b}. In such an approach, the Hilbert space of electromagnetic fields, $\mathbb{M}$, is the arena where the physical effects of light-matter interaction are modeled using the scattering operator, and where they become computationally accessible through the T-matrix due to the one-to-one connection between the scattering operator and the T-matrix. The tools for Hilbert spaces result in a compact notation with direct connections to physically relevant quantities. For example, the scalar product $\braket{f}{g}$ \cite{Gross1964} between two fields in $\mathbb{M}$, $\ket{f}$ and $\ket{g}$, can be used to express and compute the number of photons \cite{Zeldovich1965} in a given field $\ket{f}$ as $\braket{f}$, or its energy \cite[\S 9, Chap.~3]{Birula1975} as $\expval{\op{H}}{f}$, where $\op{H}$ is the energy operator. Optical forces and torques have similarly simple formulations. The convenient handling of transformations in general, and Lorentz boost in particular, results in relatively simple formulas for the interaction of light with objects moving at relativistic speeds \cite{Vavilin2024}. In the present context, this would allow one to compute thermal radiation spectra of astronomical objects that are moving with respect to the measurement devices \cite{Whittam2024}. However, we will not pursue this in here. Additionally, as we will see, the operator that determines the number of thermal photons absorbed by the object can be readily defined, and the statistically expected values for computing correlations in random fields are taken over complex scalar functions, as opposed to vectorial field quantities.

The rest of the article is organized as follows. Section~\ref{sec:method} contains the derivation of the thermal radiation of a given object at a given constant temperature from the monochromatic T-matrices of the object in a frequency range appropriate to such temperature. We obtain the expression for the density of the photon emission rate per angular wavenumber, solid angle (direction), and polarization [\Eq{eq:density}]. While we use the directional Kirchhoff law for reciprocal objects, the formalism is ready for the implementation of more nuanced theories of thermal radiation, as, for example, when the emission is assumed to occur through modes other than monochromatic plane waves \cite{Miller2017,Morino2017,CorbatonArxiv2024}. Section~\ref{sec:examples} contains three examples: A chain of dielectric spheres, a metallic helix, and a chiral molecule. The emission spectrum of the chiral BINOL molecule is best understood as the basic spectrum to be scaled by the number of molecules in a solution. The thermal radiation of the molecule shows a small imbalance in the emission of the two circular polarizations, which we quantify by the \textit{thermal g-factor}, defined in analogy to the dissymmetry g-factor in chiro-optical spectroscopy. An optimized silver helix of approximately \SI{200}{\nano\meter} in size shows a much larger thermal g-factor, with a maximum spectral value of about 1.8, when 2 is the absolute upper bound. Section~\ref{sec:conclusion} concludes the article with a brief outlook.

The computer code and the T-matrices used in this article are publicly available. The software implementation, available at \url{https://github.com/jdmazo-vasquez/TMatricesThermalRadiation}, uses the recently released \pkg{treams} Python package \cite{Beutel2023b}, which is available at \url{https://github.com/tfp-photonics/treams}, and the T-matrices are stored in a database at \url{https://tmatrix.scc.kit.edu} using a community-driven data format \cite{Asadova2025}. Conveniently, the T-matrices for composite objects made from spheres, including multilayered spheres with isotropic chiral material parameters, can be computed natively within \pkg{treams}.

\section{Theory\label{sec:method}}
Appendix~\ref{app:setting} contains a collection of expressions from the algebraic setting that are relevant here. In such a setting, the scattering operator $\op{S}$ allows one to obtain the total outgoing field $\ket{g}$ produced by the interaction between the object and any given incoming illumination $\ket{f}$. It is often written as a function of another operator, $\op{T}$:
\begin{equation}
\label{eq:sit}
	\ket{g}=\op{S}\ket{f}=\left(\Id+\op{T}\right)\ket{f}\,.
\end{equation}
The operator $\op{T}$ is essentially the T-matrix \cite{Waterman1965}, or transition matrix, which contains all information about the linear interaction of light with a given object. 

The difference between the number of incoming and outgoing photons upon interaction, that is, the absorption under illumination $\ket{f}$, can be computed as:
\begin{equation}
	\begin{split}
	&\braket{f}-\braket{g}=\bra{f}\Id-\op{S}^\dagger\op{S}\ket{f}\\
	&=\bra{f}{{-}\op{T}} - \op{T}^\dagger - \op{T}^\dagger \op{T}\ket{f}=\bra{f}\op{Q}\ket{f}\, ,
	\end{split}
\end{equation}
where $\op{X}^\dagger$ denotes the adjoint, or Hermitian conjugate, of $\op{X}$, and the last line defines the absorption operator $\op{Q}$. 

In its most common embodiment, the T-matrix maps the coefficients of an incident monochromatic field expanded in regular multipoles to the coefficients of the corresponding scattered field expanded in outgoing multipoles, which are singular at the origin.

In the basis of multipolar fields, the absorption operator $\op{Q}$ can be written as:
\begin{equation}
	\label{eq:qkk}
		\op{Q}=\sum_{jm\lambda}\sum_{\jbar\mbar\barlambda}\intdkmeasure \ \op{Q}^{jm\lambda}_{\jbar\mbar\barlambda} (k) |k j m \lambda\rangle\langle \barlambda\mbar\jbar k|\, ,
\end{equation}
where the triplets $(jm\lambda)$ and $(\jbar\mbar\barlambda)$ label the rows and columns of the monochromatic absorption matrix $\matr{{\op{Q}}}(k)$, respectively. The label $j=1,2,...$ is the multipolar degree with $j=1$ corresponding to dipoles, $j=2$ to quadrupoles, and so on, $m=-j,-j+1,\ldots, j$ is the component of the angular momentum in the $z$ direction, and $\lambda=\pm 1$ is the helicity, or polarization handedness. Appendix~\ref{app:setting} contains explicit expressions for these multipolar fields. There is a single angular wavenumber integral in \Eq{eq:qkk}, as opposed to a double integral $\int\text{d}k\int\text{d}\bar{k}$, because we assume that the operator $\op{T}$ does not couple different frequencies.

We note that the conventions for the multipolar expansions in Eq.~(\ref{eq:qkk}) are different from the typical conventions \cite{Mishchenko1996}. The changes are beneficial for the extension of the T-matrix method to a polychromatic setting \cite{Vavilin2023}. Most notably, the outgoing multipoles are multiplied by a factor of 1/2 with respect to the common definitions, which then results in the relation in \Eq{eq:sit} between $\op{S}$ and $\op{T}$, instead of the common $\matr{\op{\tilde{S}}}(k)=\matr{\op{I}}+2\matr{\op{\tilde{T}}}(k)$ in the monochromatic setting.

With the typical conventions for the monochromatic T-matrices $\matr{\op{\tilde{T}}}(k)$, also adopted in \pkg{treams}, we can obtain ${\matr{\op{Q}}}(k)$ from $\matr{\op{\tilde{T}}}(k)$ as:  
\begin{equation}
	\label{eq:qt}
	{\matr{\op{Q}}}(k)=-2{\matr{\op{\tilde{T}}}}(k)-2\matr{\op{\tilde{T}}}^\dagger(k)-4\matr{\op{\tilde{T}}}^\dagger(k)\matr{\op{\tilde{T}}}(k)\,.
\end{equation}
 
Then, the singular value decomposition (SVD) can be used at each $k$ to write:
\begin{equation}
	\label{eq:svdqk}
	\matr{{\op{Q}}}(k)=\sum_s q_s^2(k) \vec{s}(k)\vec{s}^\dagger(k)\text{, with } \vec{s}^\dagger(k)\vec{u}(k)=\delta_{su}\, ,
\end{equation}
where the singular values meet $q_s^2(k)\ge 0$, the singular vectors $\vec{s}(k)$ form an orthonormal basis for the subspace of absorbed fields at each $k$, and $\delta_{su}$ is the Kronecker delta. 

We now plug \Eq{eq:svdqk} into \Eq{eq:qkk} and obtain a compact expression for $\op{Q}$:
\begin{equation}
	\begin{split}
	\label{eq:qsk}
		\op{Q}&=\sum_{jm\lambda}\sum_{\jbar\mbar\barlambda}\intdkmeasure \\
		&\sum_s q_s^2(k) \vec{s}_{jm\lambda}(k)\vec{s}^*_{\jbar\mbar\barlambda}(k)|k j m \lambda\rangle\langle \barlambda\mbar\jbar k|\\
		&=\sum_s \intdkmeasure q_s^2(k)\ket{ks}\bra{sk}\, ,
	\end{split}
\end{equation}
where we have implicitly defined:
\begin{equation}
	\label{eq:ks}
	\ket{ks}=\sum_{jm\lambda}\vec{s}_{jm\lambda}(k)\ket{kjm\lambda}\, .
\end{equation}

In typical SVD implementations, the singular vectors are ordered by the decreasing singular values. However, we wish to ensure that the singular values and vectors vary continuously across the spectrum. Therefore, it is necessary to track the singular vectors over the angular wavenumber $k$ because crossing between the singular values may occur as $k$ changes. By assuming adiabatic variations with respect to $k$, each singular vector corresponding to an absorption matrix calculated at a discrete angular wavenumber $k_{n+1}$ can be identified with a singular mode at $k_n$ by maximizing the inner product between both sets of vectors.

It is now clear from \Eq{eq:qsk} that one can obtain the absorption upon an arbitrary illumination $\ket{f}$ as:
\begin{equation}
	\label{eq:fQf}
    \begin{split}
	\bra{f}\op{Q}\ket{f}&=\sum_s \intdkmeasure q_s^2(k)\bra{f}ks\rangle\bra{sk}f\rangle\\
    &=\sum_s \intdkmeasure q_s^2(k)|\bra{sk}f\rangle|^2\,.
    \end{split}
\end{equation}

We aim at obtaining the thermal radiation spectrum of a given object at some fixed temperature $T$. We will make the common assumption, already made by Kirchhoff (see \cite[\S~2]{Baltes1976}), that the thermal emission spectrum of an object at temperature $T$ is independent of the environment. For example, one assumes that the thermal radiation spectrum of the object kept at temperature $T$ by a warm plate is the same as if the object was immersed in and in thermal equilibrium with a Planckian radiation bath at temperature $T$. Such assumption then allows one to use the directional Kirchhoff law, which, restricted to reciprocal objects, equates the thermal emission towards direction $\phat$ at a given angular wavenumber $k$, and with polarization $\lambda$, to the absorption of the object from the bath at the same angular wavenumber and polarization, but with opposite direction $-\phat$. 

We will now use \Eq{eq:fQf} to compute the thermal emission spectrum using the directional Kirchhoff law. Treating thermal radiation requires one to introduce randomness and correlation in the formalism. We start by considering a general electromagnetic thermal bath $\ketbath$, which we will later restrict to a Planckian bath. Let us consider the plane wave expansion (see Appendix~\ref{app:setting}) of $\ketbath$:
\begin{equation}
	\label{eq:respa}
	\ketbath =\sum_{\lambda=\pm 1}\intdpinvmz \psi_\lambda(\pp)|\pp \lambda\rangle\,,
\end{equation}
where the $\psi_\lambda(\pp)$ are random variables with zero expected value, and, in general, any correlation. Here, however, we assume that the components of this thermal bath are uncorrelated in frequency, direction, and polarization, that is: 
\begin{equation}
\begin{split}
	\exval{\psi_\lambda(\pp)}&=0,\\ \exval{\psi_\lambda(\pp)\psi_{\barlambda}^*(\qq)}&=\delta_{\lambda\barlambda}\delta(\pp-\qq)\abs{\pp}^3\exval{\abs{\psi_\lambda(\pp)}^2}\,. 
\end{split}
\end{equation}
In this case, the expectation value of the absorption can be written as follows (see Appendix~\ref{app:derqpw}):
\begin{equation}
	\label{eq:qpw}
	\begin{split}		&\exval{\bra{\Psi_{\text{bath}}}\op{Q}\ket{\Psi_{\text{bath}}}}= \int_{>0}^\infty \text{d}k \int \text{d}^2\phat\\
		&\sum_{\lambda=\pm 1}k \exval{\abs{\psi_\lambda(\pp)}^2}\sum_s q_s^2(k)\abs{\sum_{jm}\sqrt{\frac{2j+1}{4\pi}} {D^j}^*_{m\lambda}(\phat)\vec{s}_{jm\lambda}(k)}^2\,,
	\end{split}
\end{equation}
where the Wigner D-matrices for spatial rotations \cite[Chapter~4]{Varshalovich1988} enter as follows
\begin{equation}
\label{eq:Dexplicit}
D_{m\lambda}^j(\phat)=D_{m\lambda}^j(\phi,\theta,0)=\exp(-\ii m \phi)d^j_{m\lambda}(\theta)\,,
\end{equation}
where $\phi=\arctantwo(k_y,k_x)$, $\theta=\arccos(k_z/k)$, $d_{m\lambda}^j(\theta)$ are the Wigner small d-matrices as defined in \cite[Chapter~4.3]{Varshalovich1988}, and $\int \text{d}^2\phat\equiv\int_{0}^\pi \text{d}\theta\sin\theta\  \int_{-\pi}^\pi \text{d}\phi$.

The second line of \Eq{eq:qpw} is the density of absorbed photons per angular wavenumber and solid angle. More precisely, in this context, the density of the photon absorption rate per angular wavenumber, solid angle, and time interval must be considered. The consideration of photon rates instead of the number of photons is necessary for treating processes of absorption and emission that occur continuously in a stationary situation such as thermal equilibrium.

Now, by the directional Kirchhoff law, the density of the photon {\em emission} rate per angular wavenumber and solid angle in $(\pp,\lambda)$ must be equal to:
\begin{equation}
	\label{eq:kdens}
	\begin{split}
		&k \exval{\abs{\psi_\lambda(-\pp)}^2}\times\\
		&\sum_s q_s^2(k)\abs{\sum_{jm}\sqrt{\frac{2j+1}{4\pi}} {D^j}^*_{m\lambda}(-\phat)\vec{s}_{jm\lambda}(k)}^2\,,
	\end{split}
\end{equation}
where $D^j_{m\lambda}(-\phat)=D^j_{m\lambda}(\phi+\pi,\pi-\theta,0)$ for the angles $(\phi,\theta)$ corresponding to the direction of $\pp$.

\begin{figure*}[ht]
    \begin{center}
    \begin{tabular}{ccc}
     \includegraphics[width=0.34\linewidth]{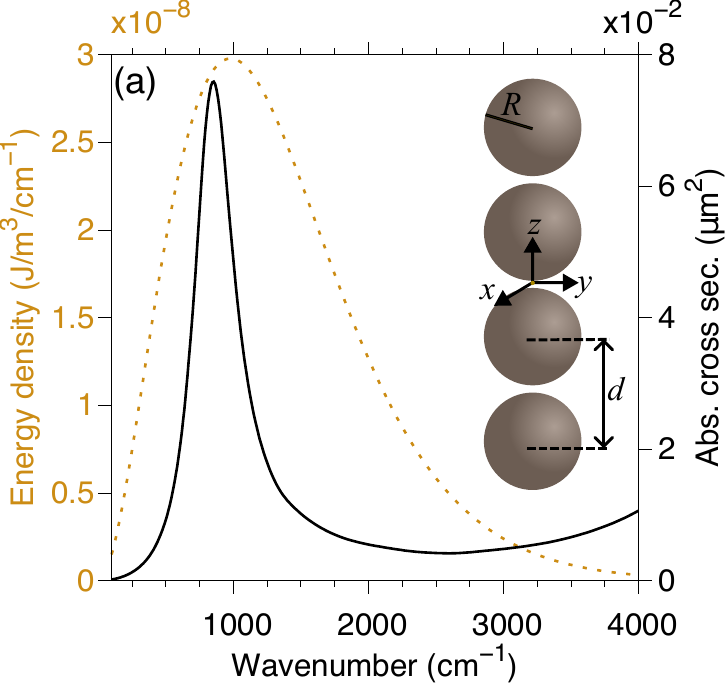}
     \includegraphics[width=0.29\linewidth]{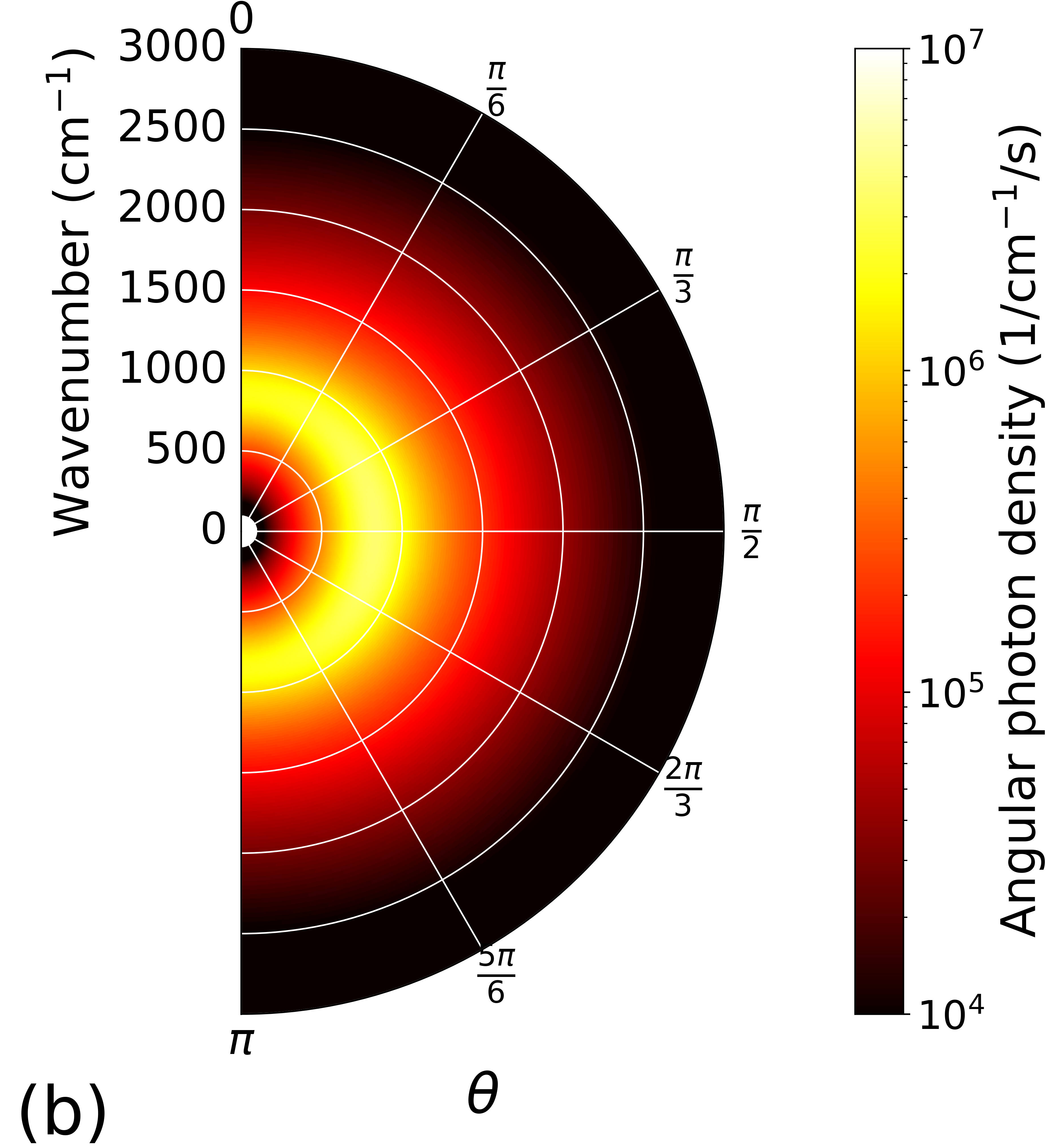}
     \includegraphics[width=0.31\linewidth]{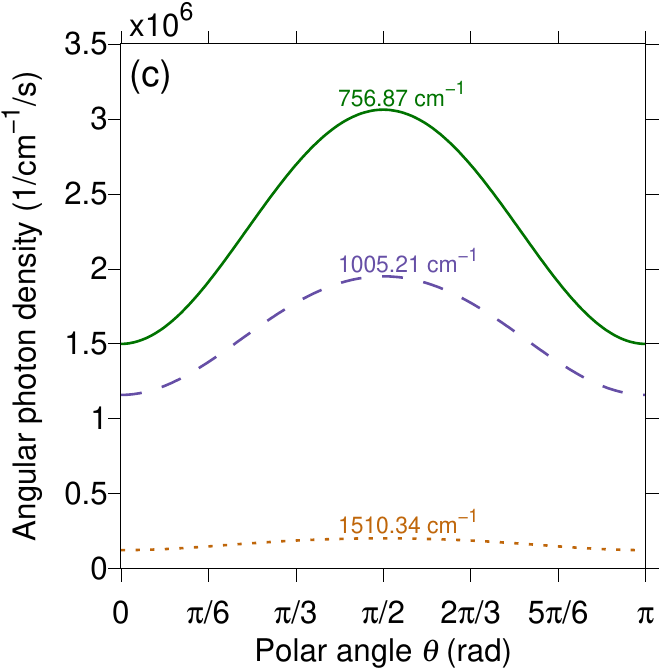}
    \end{tabular}
    \end{center}
    \caption{(a) Energy density of a Planckian thermal bath at $T = \SI{500}{\kelvin}$ and absorption cross section spectrum of the chain of SiC spheres, both as functions of the wavenumber (inverse wavelength). The spheres under consideration have a radius of $R = \SI{250}{\nano\meter}$ and the distance between the centers is $d = \SI{520}{\nano\meter}$. A schematic representation is shown in the inset figure. This structure exhibits a peak in absorption around \SI{780}{\invcm}, whereas the maximum of the energy density curve occurs approximately at \SI{980}{\invcm}. (b) Photon emission rate density as a function of the wavenumber and the polar angle $\theta$ for a fixed azimuthal angle $\phi = \pi/4$. The emission spectrum is independent of $\phi$ because the object has axial symmetry. (c) Cuts along constant wavenumbers from panel (b), solid, dashed, and dotted lines correspond to \SI{756.87}{\invcm}, \SI{1005.21}{\invcm}, and \SI{1510.34}{\invcm}, respectively.}
    \label{fig1}
\end{figure*}

That is, if we consider the plane wave decomposition of the random thermal radiation from the object $\ketobject$
\begin{equation}
\ketobject=\sum_{\lambda=\pm 1}\intdpinvmz \phi_\lambda(\pp)|\pp \lambda\rangle\, ,
\end{equation}
with 
\begin{equation}
\begin{split}
	\exval{\phi_\lambda(\pp)}&=0,\text{ and }\\
    \exval{\abs{\phi_\lambda(\pp)}^2}&=\mathbb{E}\left\{ \abs{\braket{\mathbf{k}\,\lambda}{\Phi_\text{thermal}}}^2 \right\}\, ,
    \end{split}
\end{equation}
together with the assumptions implicit in the directional Kirchhoff law, namely that the emission will be uncorrelated in $k$, direction, and polarization, the total emission rate can be written as:
\begin{equation}
	\begin{split}
		&\exval{\langle \Phi_\text{thermal} \ketobject}=\\
		&\sum_{\lambda=\pm 1}\int_{>0}^\infty \text{d}k \int \text{d}^2\phat \, k \mathbb{E}\left\{ \abs{\braket{\mathbf{k}\,\lambda}{\Phi_\text{thermal}}}^2 \right\}\,,
	\end{split}
\end{equation}
where $k\mathbb{E}\left\{ \abs{\braket{\mathbf{k}\,\lambda}{\Phi_\text{thermal}}}^2 \right\}$ is equal to the expression in (\ref{eq:kdens}).

We now particularize the illumination statistics to a Planckian bath at temperature $T$, which, besides being uncorrelated in frequency, direction, and polarization, as assumed so far, is also isotropic and unpolarized. Then (see \cite[Eq.~(B9)]{CorbatonArxiv2024}):
\begin{equation}
	\exval{\abs{\psi_\lambda(\pp)}^2}=\frac{\cz}{2\pi k \left(\exp\left(\frac{\hbar \cz k }{k_B T}\right)-1\right)}\, ,
\end{equation}
which we substitute in \Eq{eq:kdens} to obtain the final result:
\begin{equation}
	\label{eq:density}
	\boxed{
	\begin{aligned}
		&k\mathbb{E}\left\{ \abs{\braket{\mathbf{k}\,\lambda}{\Phi_\text{thermal}}}^2 \right\} =\frac{\cz}{2\pi \left(\exp\left(\frac{\hbar \cz k }{k_B T}\right)-1\right)}\times\\
		&\sum_sq_s^2(k)\abs{\sum_{jm}\sqrt{\frac{2j+1}{4\pi}} {D^j}^*_{m\lambda}(\phi+\pi,\pi-\theta,0)\vec{s}_{jm\lambda}(k)}^2
	\end{aligned}
	}
	\,.
\end{equation}

Therefore, one can obtain all the information about the thermal spectrum of a given object when the monochromatic T-matrices $\matr{\op{\tilde{T}}}(k)$ in a range of $k$ appropriate to the temperature are available. The computations involve the SVD of the $\matr{\op{Q}}(k)$ in \Eq{eq:qt} at each frequency, tracking the modes over $k$, and using \Eq{eq:density}. We have verified that the calculations of the thermal spectrum for a single homogeneous sphere coincide with the results obtained semi-analytically using the Mie coefficients of the sphere and the directional Kirchhoff law.

It is possible to avoid the SVD by taking a different computational strategy where the multipolar decomposition of monochromatic plane waves is used together with the monochromatic absorption matrices $\matr{{\op{Q}}}(k)$. We have chosen the presented strategy with an eye on the future implementation of more nuanced theories of thermal radiation beyond the directional Kirchhoff law. Thermal radiation effects are currently considered in a variety of systems, such as for example graphene patches and time-variying media \cite{Manjavacas2014,Yu2017,Miller2015,Greffet2018,Vazquez2023,Eriksen2025}, which constitute a challenge for many of the existing theories. For example, the directional Kirchhoff law assumes that, for all objects, monochromatic plane waves are the modes of light into which energy is independently absorbed and emitted. Recent work \cite{Miller2017} postulates that emission and absorption occur through monochromatic modes that depend on the object. Such modes are obtained from the SVD of the S-matrix. The postulate leads to directional correlations in the thermal radiation that are absent in the directional Kirchhoff law. In \cite{Morino2017}, the authors successfully analyze measurements of thermal radiation using the modes of open spherical cavities. Such modes, characterized by complex frequencies, can be understood as polychromatic modes with Lorentzian-like lineshapes. As seen in \cite[Fig.~3]{Morino2017}, the results of such a theory differ from those of the typical monochromatic approach. The formalism developed in this article is a basis for similar extensions. For example, the $\ket{k s}$ and the $q_s^2(k)$ are the building blocks of a polychromatic theory of emission \cite{CorbatonArxiv2024}, which, contrary to the monochromatic ones, permits different rates of emitted and absorbed photons, and can be applied to both thermal emission and luminescence.

The following formulas for the common rotationally-averaged absorption cross sections will be used in the next section: 
\begin{equation}
	\label{eq:xsecs}
	\begin{split}
		\sigma^{\text{abs}}(k)&=\frac{\pi}{2 k^2}\sum_sq_s^2(k)=\frac{\pi}{2k^2}\text{Trace}\left\{\matr{\op{Q}}(k)\right\}\,,\\
		\sigma^{\text{abs}}(k)&=\sigma^{\text{abs}}_{\lambda=+1}(k)+\sigma^{\text{abs}}_{\lambda=-1}(k)\text{\, , and}\\
		\sigma^{\text{abs}}_{\lambda}(k)&=\frac{\pi}{2 k^2}\sum_sq_s^2(k)\sum_{jm}\abs{\vec{s}_{jm\lambda}(k)}^2\, .
	\end{split}
\end{equation}
Their derivation can be found in Appendix~\ref{app:derqpw}.

At this point, it is worth mentioning one of the current limitations of the T-matrix method, which concerns the extreme near field. Namely, it is unclear whether the standard algorithms apply when computing the interactions between two objects that penetrate each other's smallest circumscribed spheres. This issue is under investigation, and there are some solutions \cite{Theobald2017,Egel2017,Martin2019,Schebarchov2019,Rother2021,barkhan2022,Lamprianidis2023}.

\section{Examples\label{sec:examples}}
This section contains three examples: A chain of dielectric spheres, a metallic helix, and a chiral molecule. Since defining a temperature for a single molecule is not straightforward, the emission spectrum of the chiral molecule is best understood as the basic spectrum to be scaled by the number of molecules in a solution, for example. It should be noted that the molecular T-matrices are those for the ground state, and a more accurate computation at the considered temperature should also take into account T-matrices for the excited vibrational states, and non-coherently combine the ground and excited state spectra with populations according to the temperature. Similarly, the material parameters used for computing the T-matrices of the spheres and the helix are those corresponding to room temperature, and a more accurate computation should use material parameters for the desired temperature.

\subsection{Chain of dielectric spheres}
Let us consider first a chain made of four spheres, each with \SI{250}{\nano\meter} radius, and separated by a distance $d = \SI{520}{\nano\meter}$, as illustrated in the inset of Fig.~\ref{fig1}. The spheres are made of SiC, and the structure has a resonance around \SI{780}{\invcm}. Data for the SiC permittivity were taken from \cite{Larruquert2011, Polyanskiy2024}. The absorption cross section of the chain, and the energy density of the Planckian thermal bath, at temperature $T = \SI{500}{\K}$, are shown in Fig.~\ref{fig1}(a) as functions of the wavenumber (inverse wavelength) in \si{\invcm}. For computing the T-matrix of the structure, we used the package \pkg{treams}, which has native support for the computation of composite objects made out of spheres. The maximum value for the multipolar degree was chosen to be $j_{\text{max}} =10$. The photon emission rate density is shown in Fig.~\ref{fig1}(b) as a function of the wavenumber and the polar angle $\theta$, for a fixed azimuthal angle $\phi = \pi/4$. The chain of spheres emits mostly in the region between \num{700} and \SI{900}{\invcm} as a consequence of its absorption peak and the maximum of the spectrum of the Planckian thermal bath. The small spacing of the spheres in the chain leads to a dipole-shaped emission spectrum as the emission maximum occurs close to $\theta = \pi/2$, as shown in Fig.~\ref{fig1}(c). The emission does not depend on the azimuthal angle $\phi$ because the structure is symmetric under rotations around the $z$-axis. The optical properties of the material composing the chain do not exhibit chiral features and, therefore, emission occurs similarly for both helicities $\lambda = \pm 1$. This is sharply different in the following example.

\subsection{Metallic helix}\label{subsec:helix}
\begin{figure}[h!]
    \includegraphics[width=\linewidth]{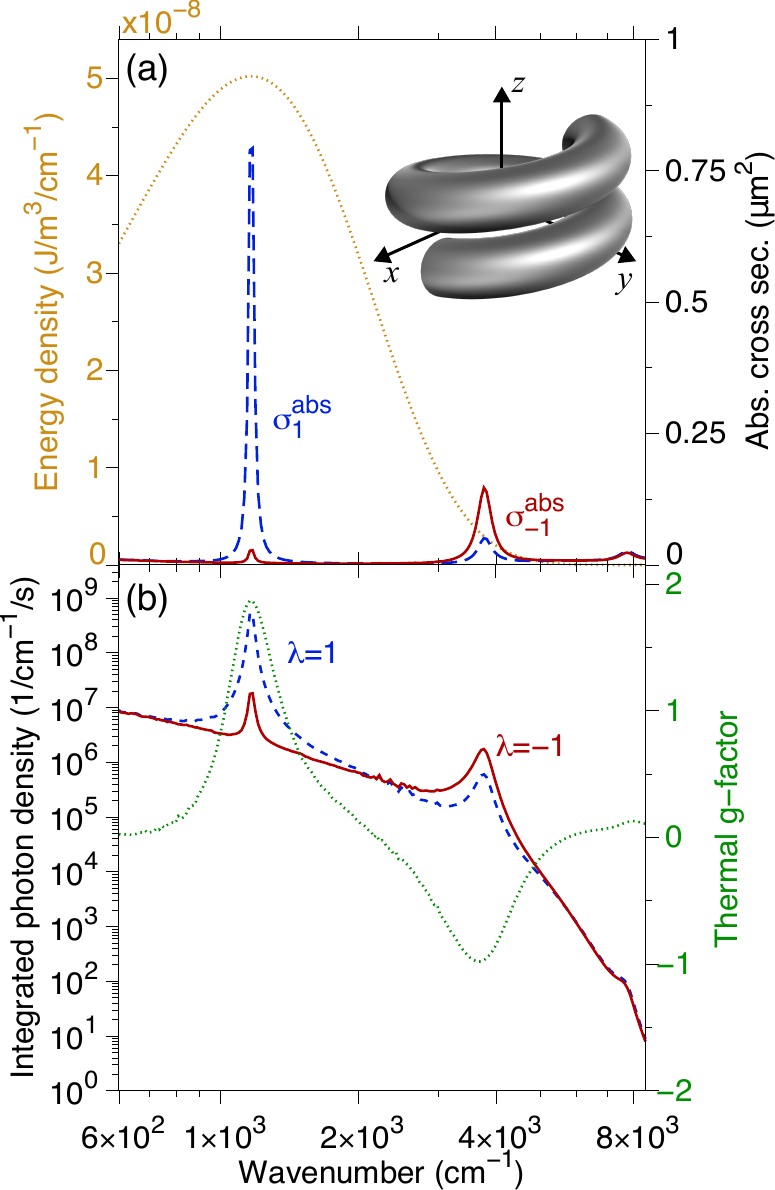}
	\caption{(a) Energy density of a Planckian thermal bath at $T = \SI{595}{\kelvin}$ and helicity-dependent rotationally-averaged absorption cross sections of a silver helix with strong em-chirality around $\SI{35}{\tera\hertz}$, both as functions of the wavenumber (inverse wavelength). (b) Integrated photon emission rate density for the helix and the corresponding thermal dissymmetry g-factor, defined in \Eq{eq:gth}, as functions of the wavenumber. False color plots for the angular emission densities for both helicities are included in Appendix~\ref{app:helixplots}. }
    \label{fig2}
\end{figure}

We now consider a silver helix, which is an example of a chiral object. The thermal radiation from chiral structures is interesting in particular because of its potential imbalance between the two circular polarizations \cite{Lee2007,Khandekar2019,Dyakov2018,Enders2024}.  

The helix parameters have been chosen to yield a strong chiral optical response, that is, to maximize its differential response to left- and right-hand polarized illumination. The geometric parameters are: A helix radius of $\SI{251.6}{\nano\meter}$ and a pitch of $\SI{168.6}{\nano\meter}$ while the (right-handed) helix describes a total number of $\num{1.407}$ turns. The wound-up cylindrical silver wire that constitutes the helix has a radius of $\SI{73.32}{\nano\meter}$ and terminates in hemispherical end caps. The chiral response is quantified as \emph{electromagnetic chirality} (em-chirality) \cite{FerCor2016}. The helix reaches a maximal normalized em-chirality of 92.4\% when the frequency is equal to $\SI{34.78}{\tera\hertz}$.

The T-matrices for the helix were obtained from finite element method (FEM) simulations using the software \pkg{JCMsuite}. In the scattering simulation, the helix is illuminated with a multipole of unit amplitude. In a postprocessing step, the scattered field is expanded into multipoles, truncated at a multipolar degree of ${j_{\textnormal{max}} = 5}$. The expansion coefficients correspond to the entries of one column of the T-matrix. Repeating for other illuminating multipoles up to $j_{\textnormal{max}}$, the full T-matrix is obtained. The process is repeated for different frequencies. Frequency-dependent material parameters for silver are interpolated from data in Ref.~\cite{Hagemann_Optical_1975}, as provided at \url{https://refractiveindex.info}. Ambient material parameters correspond to vacuum. 

The helix is immersed in a Planckian thermal bath at \SI{595}{\K}, such that the maximum of the bath's energy density coincides with the maximum absorption of the helix, as shown in Fig.~\ref{fig2}(a). For this particular object, the strong em-chirality in the form of a significant imbalance between the absorption of the two helicities, as seen in Fig.~\ref{fig2}(b), leads to different emission rates depending on the helicity of light. Here, we quantify such imbalance by integrating the angular photon emission rate density over the whole solid angle, hence obtaining the photon emission rate density per wavenumber:
\begin{equation}
	\mathcal{I}_\lambda (k)  = \int \text{d}\mathbf{\hat{k}}\, k \mathbb{E}\{ \abs{\braket{\mathbf{k}\,\lambda}{\Phi_\text{thermal}}}^2\}\, , 
\end{equation}
such that the total number of photons of each helicity emitted per second is 
\begin{equation}\label{eq11}
	N_\lambda = \int_{>0}^\infty \text{d}k \, \mathcal{I}_\lambda (k)\, .
\end{equation}
The imbalance between the rates of photon emission for both helicities can be computed as 
\begin{equation}
	\label{eq:gth}
	g_{\text{th}} (k) = 2 \frac{\mathcal{I}_{+1}(k) - \mathcal{I}_{-1}(k)}{\mathcal{I}_{+1} (k) + \mathcal{I}_{-1} (k)}\, ,
 \end{equation}
 here defined as the \textit{thermal g-factor}, in analogy to the definition of the dissymmetry g-factor in chiro-optical spectroscopy.

Figure~\ref{fig2}(b) shows the integrated photon emission rate densities for the chiral helix for each helicity, and the thermal g-factor, both as functions of the wavenumber. The positive g-factor close to \SI{e3}{\invcm} indicates a higher emission of photons with positive helicity, in contrast with the negative g-factor around \SI{4e3}{\invcm}, where more photons with negative helicity are emitted. After computing the integral over the wavenumber in Eq.~\eqref{eq11}, one obtains $N_{+1} = \num{3.53e10}$ and $N_{-1} = \num{5.91e9}$ emitted photons per second with positive and negative helicity, respectively. Such a difference should be straightforward to observe experimentally: The difference between positive and negative helicity emission power is on the order of the emission power itself, and there has been prior work on measuring the thermal spectra of single micro-objects \cite{Fenollosa2019,Hamdan2022,Fenollosa2025}.

Noise-like fluctuations can be seen in some frequency regions in Fig.~\ref{fig2}(b). These are due to numerical noise in the calculated T-matrix entries. Such numerical noise also determines the lower bound on the frequency range, as follows. As the wavelength of light grows, the interaction with the helix decreases, and the entries of the T-matrix become smaller. At some point, the ratio between the absolute value of such entries and the numerical noise reaches an unacceptable level.

\subsection{Chiral molecules}
The method proposed in this paper is general and can also be applied to compute the thermal emission spectrum of molecules. Since chiral molecules absorb the two helicities of light with different efficiencies, we expect a non-zero $g_{\text{th}}$ in their thermal spectrum. 

\begin{figure}
    \includegraphics[width=\linewidth]{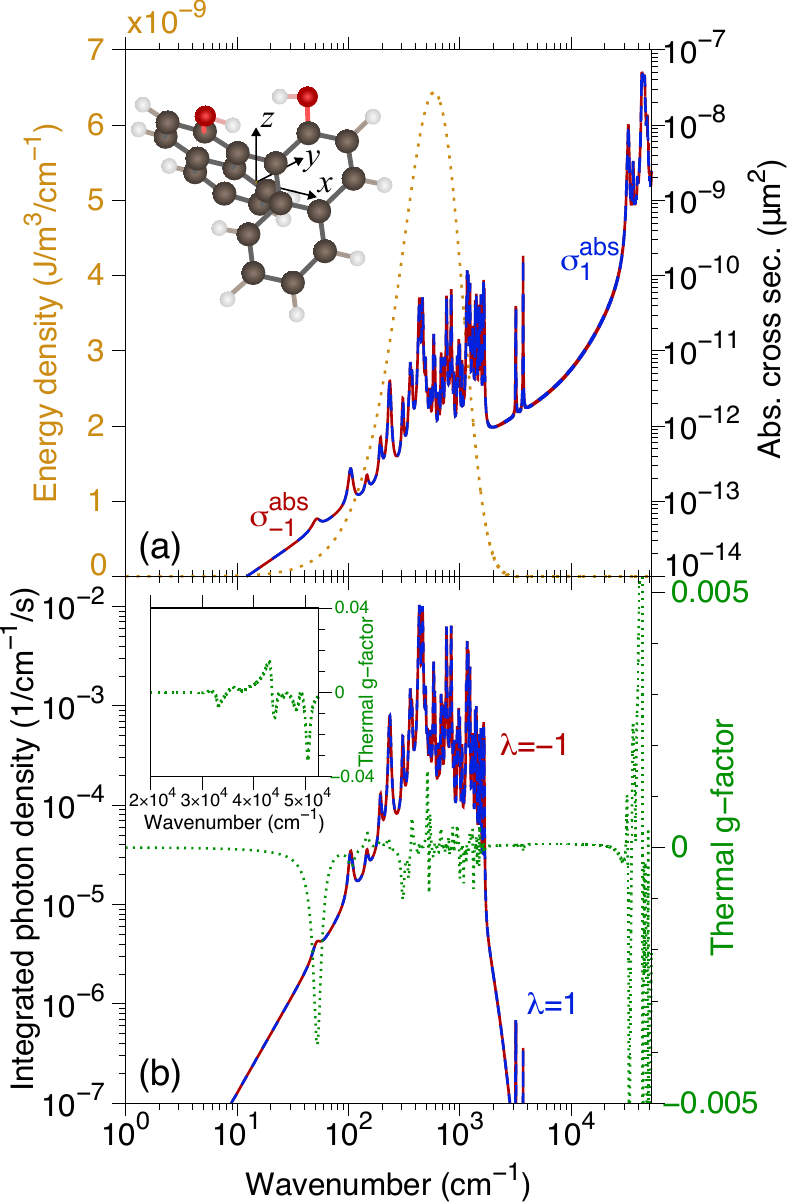}
	\caption{(a) Energy density of a Planckian thermal bath at $T = \SI{300}{\kelvin}$ and helicity-dependent orientationally-averaged absorption cross sections of a R-BINOL molecule, depicted in the inset. The molecule is surrounded by Chloroform with electric permittivity $\varepsilon = 4.711$. (b) Integrated photon emission rate density for the R-BINOL molecule and the corresponding thermal dissymmetry g-factor, defined in \Eq{eq:gth}, as functions of the wavenumber.}
    \label{fig3}
\end{figure}

As a final example, we present here a 1,1'-Bi-2-naphthol (BINOL) molecule solvated in Chloroform, composed of 20 carbon atoms, 14 hydrogen atoms, and 2 oxygen atoms, arranged in two connected naphthalene subunits having one alcohol (-OH) group each, as depicted in the inset of Fig.~\ref{fig3}(a). Here, we consider only one of the two enantiomers, namely the R-enantiomer of the BINOL. The T-matrices of molecules are obtained by quantum-chemical calculations of the molecular polarizabilities, which are mapped onto T-matrices of dipolar degree \cite{FerCor2018}. The dynamic polarizabilities are used to construct T-matrices separately for vibrational and electronic excitations. The final result is obtained by the frequency-wise sum of these two kinds of T-matrices. Appendix~\ref{app:quantumchem} contains details on the approach that we used, which is based on density functional theory (DFT) and its time-dependent variant (TD-DFT) for the optical response. Figure~\ref{fig3}(a) shows the helicity-dependent absorption cross sections of one single molecule of R-BINOL immersed in a Planckian thermal bath at \SI{300}{\K}, and the energy density of the bath, both as functions of the wavenumber. Absorption peaks are exhibited in the region where vibrational transitions occur (between \SI{e2}{\invcm} and \SI{1.5e3}{\invcm}), as well as where electronic transitions occur (above \SI{e4}{\invcm}). The maximum energy density of the Planckian thermal bath occurs at \SI{588.26}{\invcm}, suggesting that energy absorption by the molecule leads to transitions in vibrational states.

Figure~\ref{fig3}(b) depicts the integrated photon rate density emitted by the molecule and the corresponding $g_{\text{th}}(k)$, both as functions of the wavenumber. Overall, the absolute value of the $g_{\text{th}}(k)$ factors is rather small, except for a peak around \SI{50}{\invcm} indicating higher emission of photons with negative helicity, and other peaks in the near UV region. The relative emission is, however, rather small in both cases. The total emitted photons rates for two helicities are $N_{+1} = \SI{1.116001}{\second\tothe{-1}}$ and $N_{-1} = \SI{1.115987}{\second\tothe{-1}}$. Since thermal radiation is not coherent, one can obtain the spectra for a given number of molecules by simply multiplying the spectra of a single molecule by said number. As a reference, one cubic centimeter of pure liquid BINOL contains $\sim \num{2.73e21}$ molecules. 

\section{Conclusion and outlook\label{sec:conclusion}}
This work provides a basic formalism for computing thermal radiation by combining Waterman's T-matrix method with an algebraic approach to light-matter interactions. Given the respective T-matrices, the formalism can be applied to objects such as micro- and nano-particles, clusters thereof, an also to molecules, allowing one to study the thermal radiation of molecular solutions. There are several directions towards which the formalism can be extended. We mention some of them to conclude the article. 

The setting introduced here is ready for implementing more nuanced thermal emission theories beyond the currently implemented directional Kirchhoff law. In such theories, the basic modes for emission and absorption of thermal energy are not monochromatic plane waves. 

Existing formulas for changing between inertial reference frames \cite{Vavilin2024}, developed within the algebraic formalism, would allow one to obtain the thermal spectra of relativistically moving objects in a straightforward way.

Finally, quantities relevant to near-field radiative heat transfer, such as the transfer between two given objects within a larger cluster, can be readily and accurately computed as long as the two objects do not invade each other's circumscribed sphere.

\section*{Acknowledgements}
J.D.M.-V. is part of the Max Planck School of Photonics, supported by the German Federal Ministry of Education and Research (BMBF), the Max Planck Society, and the Fraunhofer Society.
M.K. and C.R. acknowledge support by the Deutsche Forschungsgemeinschaft (DFG, German Research Foundation) under Germany’s Excellence Strategy via the Excellence Cluster 3D Matter Made to Order (EXC-2082/1-390761711) and from the Carl Zeiss Foundation via the CZF-Focus@HEiKA Program. M.K. and C.R. acknowledge funding by the Volkswagen Foundation. I.F.C. and C.R. acknowledge support by the Helmholtz Association via the Helmholtz program “Materials Systems Engineering” (MSE). 
L.R. acknowledges support by the Karlsruhe School of Optics \& Photonics (KSOP).
L.R., I.F.C., and C.R. acknowledge support by the Deutsche Forschungsgemeinschaft (DFG, German Research Foundation) -- Project-ID 258734477 -- SFB 1173.
M.N. and C.R. acknowledge support by the KIT through the “Virtual Materials Design” (VIRTMAT) project. 
The authors are grateful to the company JCMwave for their free provision of the FEM Maxwell solver JCMsuite.
M.K. and C.R. acknowledge support by the state of Baden--Württemberg through bwHPC and the German Research Foundation (DFG) through grant no. INST 40/575-1 FUGG (JUSTUS 2 cluster) and the HoreKa supercomputer funded by the Ministry of Science, Research and the Arts Baden--Württemberg and by the Federal Ministry of Education and Research. 

\appendix
\section{Relevant expressions from the algebraic setting\label{app:setting}}
The electric field of a particular solution $\ket{f}$ is expanded into plane waves of well-defined helicity $\ket{\pp \lambda}$ as:
\begin{equation}
	\label{eq:xpans}
	 \Ert \equiv \sum_{\lambda=\pm1 }\int_{\mathbb{R}^3-\zerovec} \frac{\text{d}^3 \pp}{k} \, \ff{\lambda} \, \ket{\pp\lambda},
\end{equation}
and the plane waves are defined as:
\begin{equation}
	\begin{split}
		&\ket{\pp \lambda}\equiv\\
		&\sqrt{\frac{\cz\hbar}{ \epsz}}\, \frac{1}{\sqrt{2}} \frac{1}{\sqrt{(2\pi)^3}}\, k \, \mathbf{\hat{e}}_\lambda({\phat}) \exp(- \ii k\cz t ) \exp(\ii \pp \cdot \rr)\, ,
	\end{split}
\end{equation}
where $\cz$ is the speed of light in vacuum, $\hbar$ the reduced Planck constant, $\epsz$ the permittivity in vacuum, $\pp$ is the wavevector, and $\lambda$ the helicity, or polarization handedness.

The expansion in multipoles of well-defined helicity reads:
\begin{equation}
	\label{eq:ypans}
	\begin{split}
		&{\Ert}^{\text{reg/in/out}} \equiv\\
		&\int_{>0}^\infty \text{d}k \, k \, \sum_{\lambda=\pm 1} \sum_{j=1}^{\infty} \sum_{m=-j}^j \, \FF{\lambda} \, \ket{k j m \lambda}^{\text{reg/in/out}}\, ,
	\end{split}
\end{equation}
and the regular, incoming, and outgoing multipoles $\ket{k j m \lambda}^{\text{reg/in/out}}$ are defined as:

\begin{widetext}
	\begin{equation}
		\begin{split}
		\label{eq:mpdef}
			&\ket{kjm\lambda}^{\text{reg}}\equiv\mathbf{S}^\text{reg}_{jm\lambda}(k,\rr,t)=\\
			&- \sqrt{\frac{\cz\hbar}{\epsz}} \frac{1}{\sqrt{2\pi}} \, k \, \ii^j  \times\Big(  \exp(-\ii k \cz t)\, \mathbf{N}^{\text{reg}}_{jm}(k|\rr|, \hat{\rr}) + \lambda \,\exp(-\ii k\cz t) \,  \mathbf{M}^{\text{reg}}_{jm}(k|\rr|, \rhat ) \Big)\, ,\\
			&\ket{kjm\lambda}^{\text{in/out}}\equiv\mathbf{S}^\text{in/out}_{jm\lambda}(k,\rr,t)=\\
			&- \frac{1}{2} \sqrt{\frac{\cz\hbar}{\epsz}} \frac{1}{\sqrt{2\pi}} \, k \, \ii^j\times \Big(  \exp(-\ii k \cz t)\, \mathbf{N}^{\inout}_{jm}(k|\rr|, \hat{\rr}) + \lambda \,\exp(-\ii k\cz t) \,  \mathbf{M}^{\inout}_{jm}(k|\rr|, \rhat ) \Big)\, ,
		\end{split}
	\end{equation}
\end{widetext}
where the $\mathbf{M}$ and $\mathbf{N}$ have the usual definitions (see e.g. \cite[Eqs.~(50,51)]{Vavilin2023}).

The scalar product between two fields in the Hilbert space of solutions of Maxwell equations \cite{Gross1964}, $\mathbb{M}$, can be computed as:
\begin{equation}
	\label{eq:lmsp}
	\langle f|g\rangle = \sum_{\lambda=\pm1} \int_{\mathbb{R}^3-\zerovec} \frac{\text{d}^3 \pp}{k} \, \ffstar{\lambda}\mathrm{g}_\lambda (\pp)\, ,
\end{equation}
and $\ff{\lambda}$ and $\mathrm{g}_\lambda (\pp)$ are the coefficient functions of the plane wave expansions of $\ket{f}$ and $\ket{g}$, respectively. The scalar product can also be computed as
\begin{equation}
	\label{eq:amsp}
	\langle f|g\rangle = \sum_{\lambda=\pm 1} \int_{>0}^{\infty} \text{d}k\, k \sum_{j=1}^{\infty} \sum_{m=-j}^j \, \FFstar{\lambda} \mathrm{g}_{jm\lambda}(k)\, ,
\end{equation}
where the $\FF{\lambda}$ and $\mathrm{g}_{jm\lambda}(k)$ are the coefficient functions of the expansions in multipolar fields, also known as spherical waves. Equations~(\ref{eq:lmsp}) and (\ref{eq:amsp}) are written using the conventions in \cite{Vavilin2023}, which we have included here for reference.

\section{Derivations of \Eq{eq:qpw} and \Eq{eq:xsecs}\label{app:derqpw}}
We first derive \Eq{eq:qpw}, starting from \Eq{eq:fQf}, from where it is readily seen that:
\begin{equation}
	\label{eq:first}
	\exval{\bra{f}\op{Q}\ket{f}}=\sum_s \intdkmeasure q_s^2(k)\exval{\abs{\bra{sk}f\rangle}^2}\, ,
\end{equation}
so we start with the terms $\exval{\abs{\bra{sk}f\rangle}^2}$ by using the expansion of the $\ket{ks}$ into plane waves \cite[Sec.~8.4.1]{Tung1985}:
\begin{equation}
	\ket{ks}=\sum_{jm\lambda}\vec{s}_{jm\lambda}(k)\int \text{d}^2\phat\,\sqrt{\frac{2j+1}{4\pi}}{D^j}^*_{m\lambda}(\phat)\ket{k\phat\ \lambda}\, ,
\end{equation}
with which, using that $\bra{\lambda\pp}f\rangle=f_\lambda(\pp)$, we can write:
{\small
\begin{equation}
	\label{eq:dint}
	\begin{split}
		&\exval{\abs{\bra{sk}f\rangle}^2}=\sum_{jm\lambda}\sum_{\jbar\mbar\barlambda}\vec{s}_{jm\lambda}(k)\vec{s}_{\jbar\mbar\barlambda}(k)\int \text{d}^2\ahat\int \text{d}^2\bhat\\
		& \sqrt{\frac{2j+1}{4\pi}}{D^j}^*_{m\lambda}(\ahat)\sqrt{\frac{2\jbar+1}{4\pi}}D^{\jbar}_{\mbar\barlambda}(\bhat)\exval{f_\lambda(k\ahat)f^*_{\barlambda}(k\bhat)}\, .
	\end{split}
\end{equation}
}
We now assume that the random illumination is uncorrelated in direction and polarization, that is:
\begin{equation}
	\exval{f_\lambda(k\ahat)f^*_{\barlambda}(k\bhat)}=\delta(\ahat-\bhat)\delta_{\lambda\barlambda}\exval{\abs{f_\lambda(k\ahat)}^2}\, .
\end{equation}
Under such assumption, \Eq{eq:dint} becomes:
\begin{equation}
	\label{eq:sint}
	\begin{split}
		&\exval{\abs{\bra{f}ks\rangle}^2}=\sum_{jm}\sum_{\jbar\mbar}\sum_\lambda\vec{s}_{jm\lambda}(k)\vec{s}_{\jbar\mbar\lambda}(k)\\
		&\int \text{d}^2\ahat\,\sqrt{\frac{2j+1}{4\pi}}{D^j}^*_{m\lambda}(\ahat)\sqrt{\frac{2\jbar+1}{4\pi}}D^{\jbar}_{\mbar\lambda}(\ahat)\exval{\abs{f_\lambda(k\ahat)}^2}=\\
		&\sum_\lambda\int \text{d}^2\ahat \, \abs{ \sum_{jm}\sqrt{\frac{2j+1}{4\pi}}{D^j}^*_{m\lambda}(\ahat)\vec{s}_{jm\lambda}(k)}^2\exval{\abs{f_\lambda(k\ahat)}^2}\, .
	\end{split}
\end{equation}
After substituting the last line of \Eq{eq:sint} into \Eq{eq:first}, we arrive at \Eq{eq:qpw} of the main text:
\begin{equation}
	\begin{split}
		&\exval{\bra{f}\op{Q}\ket{f}}=\sum_{\lambda=\pm 1}\int_{>0}^\infty \text{d}k \int \text{d}^2\phat\\\
		&k \exval{\abs{f_\lambda(\pp)}^2}\sum_s q_s^2(k)\abs{\sum_{jm}\sqrt{\frac{2j+1}{4\pi}} {D^j}^*_{m\lambda}(\phat)\vec{s}_{jm\lambda}(k)}^2\, .
	\end{split}
\end{equation}

To derive the expressions in (\ref{eq:xsecs}), we now go back to \Eq{eq:sint} and further assume isotropy, as in the Planckian illumination, that is: $\exval{\abs{f_\lambda(k\ahat)}^2}=\exval{\abs{f_\lambda(k)}^2}$. This allows one to solve the angular integral: 
\begin{equation}
	\label{eq:long}
	\begin{split}
		&\exval{\abs{\bra{f}ks\rangle}^2}=\sum_{jm}\sum_{\jbar\mbar}\sum_\lambda\vec{s}_{jm\lambda}(k)\vec{s}_{\jbar\mbar\lambda}(k)\exval{\abs{f_\lambda(k)}^2}\\
		&\int \text{d}^2\ahat\,\sqrt{\frac{2j+1}{4\pi}}{D^j}^*_{m\lambda}(\ahat)\sqrt{\frac{2\jbar+1}{4\pi}}D^{\jbar}_{\mbar\lambda}(\ahat)\\
		&=\sum_\lambda\exval{\abs{f_\lambda(k)}^2}\sum_{jm}\abs{\vec{s}_{jm\lambda}(k)}^2\, ,
	\end{split}
\end{equation}
where the last equality follows from solving the angular integral by substituting $D_{m\lambda}^j(\phat)=\exp(-\ii m\phi)d^j_{m\lambda}(\theta)$, solving the integral in $\phi$, and using the orthogonality properties of the real-valued small Wigner d-matrices \cite[Equation~8.3-2]{Tung1985}:
\begin{equation}
	\label{eq:sp}
	\begin{split}
		&\int_{-\pi}^\pi \text{d}\phi\int_{0}^\pi \text{d}\theta\,\sin\theta D_{\bar{m}\lambda}^{\bar{j}}(\phi,\theta,0){D^j}^*_{m\lambda}(\phi,\theta,0)\\
		&=\int_{-\pi}^\pi \text{d}\phi \, \exp\left(\ii\left(m-\bar{m}\right)\phi\right)\int_{0}^\pi d\theta\,\sin\theta\ {d}_{\bar{m}\lambda}^{\bar{j}}(\theta)d^j_{m \lambda}(\theta)\\
		&=2\pi \delta_{\bar{m}m} \int_{0}^\pi d\theta\,\sin\theta\ {d}_{\bar{m}\lambda}^{\bar{j}}(\theta){d}^j_{m \lambda}(\theta)\\
		&=\frac{4\pi}{2j+1}\delta_{\bar{m}m}\delta_{\bar{j}j}\, . 	
	\end{split}
\end{equation}
We substitute the last line of \Eq{eq:long} in \Eq{eq:first}, and also the Planckian spectrum $\exval{\abs{f_\lambda(k)}^2}=\frac{\cz}{2\pi k\left(\exp\left(\frac{\hbar \cz k }{k_B T}\right)-1\right)}$ (see \cite[Eq.~(20)]{CorbatonArxiv2024}):
\begin{equation}
	\begin{split}
		\exval{\bra{f}\op{Q}\ket{f}}&=\int_{>0}^\infty \text{d}k\,\frac{\cz}{2\pi \left(\exp\left(\frac{\hbar \cz k }{k_B T}\right)-1\right)}\\
		&\sum_s q_s^2(k)\sum_{jm\lambda}\abs{\vec{s}_{jm\lambda}(k)}^2\, ,
	\end{split}
\end{equation}
and compare the result with a different expression for the absorption under a Planckian bath, involving the absorption cross section $\sigma^{\text{abs}}(k)$:
\begin{equation}
	\exval{\bra{f}\op{Q}\ket{f}}=\int_{>0}^\infty \text{d}k\,\frac{\cz k^2}{\pi^2}\frac{1}{\left(\exp\left(\frac{\hbar \cz k }{k_B T}\right)-1\right)}\sigma^{\text{abs}}(k)\, ,
\end{equation}
which is obtained by dividing the integrand of the absorbed power in \cite[Eq.~(B4)]{CorbatonArxiv2024} by the energy $\hbar\cz k$, and hence going from Joules per second to photons per second.

Such comparison readily yields the first equation in (\ref{eq:xsecs}):
\begin{equation}
	\begin{split}
		\sigma^{\text{abs}}(k)&=\frac{\pi}{2k^2}\left[\sum_s q_s^2(k)\sum_{jm\lambda}\abs{\vec{s}_{jm\lambda}(k)}^2\right]\\
		&=\frac{\pi}{2k^2}\left(\sum_s q_s^2(k)\right)=\frac{\pi}{2k^2}\text{Trace}\left\{\matr{\op{Q}}(k)\right\}\, ,
	\end{split}
\end{equation}
where the second equality follows because the norm of $\vec{s}(k)$ is equal to one, and the second from the SVD of $\matr{\op{Q}}(k)$. The other two equations in (\ref{eq:xsecs}) are obtained by splitting the sum of the two helicities into two terms:
\begin{equation}
	\begin{split}
		\sigma^{\text{abs}}(k)&=\frac{\pi}{2k^2}\left[\sum_s q_s^2(k)\sum_{jm\lambda=+1}\abs{\vec{s}_{jm1}(k)}^2\right]\\
		&+\frac{\pi}{2k^2}\left[\sum_s q_s^2(k)\sum_{jm\lambda=-1}\abs{\vec{s}_{jm-1}(k)}^2\right]\\
		&=\sigma^{\text{abs}}_{\lambda=+1}(k)+\sigma^{\text{abs}}_{\lambda=-1}(k)\, .
	\end{split}
\end{equation}

\begin{figure*}
\begin{center}
\begin{tabular}{cc}
\includegraphics[width=0.4\linewidth]{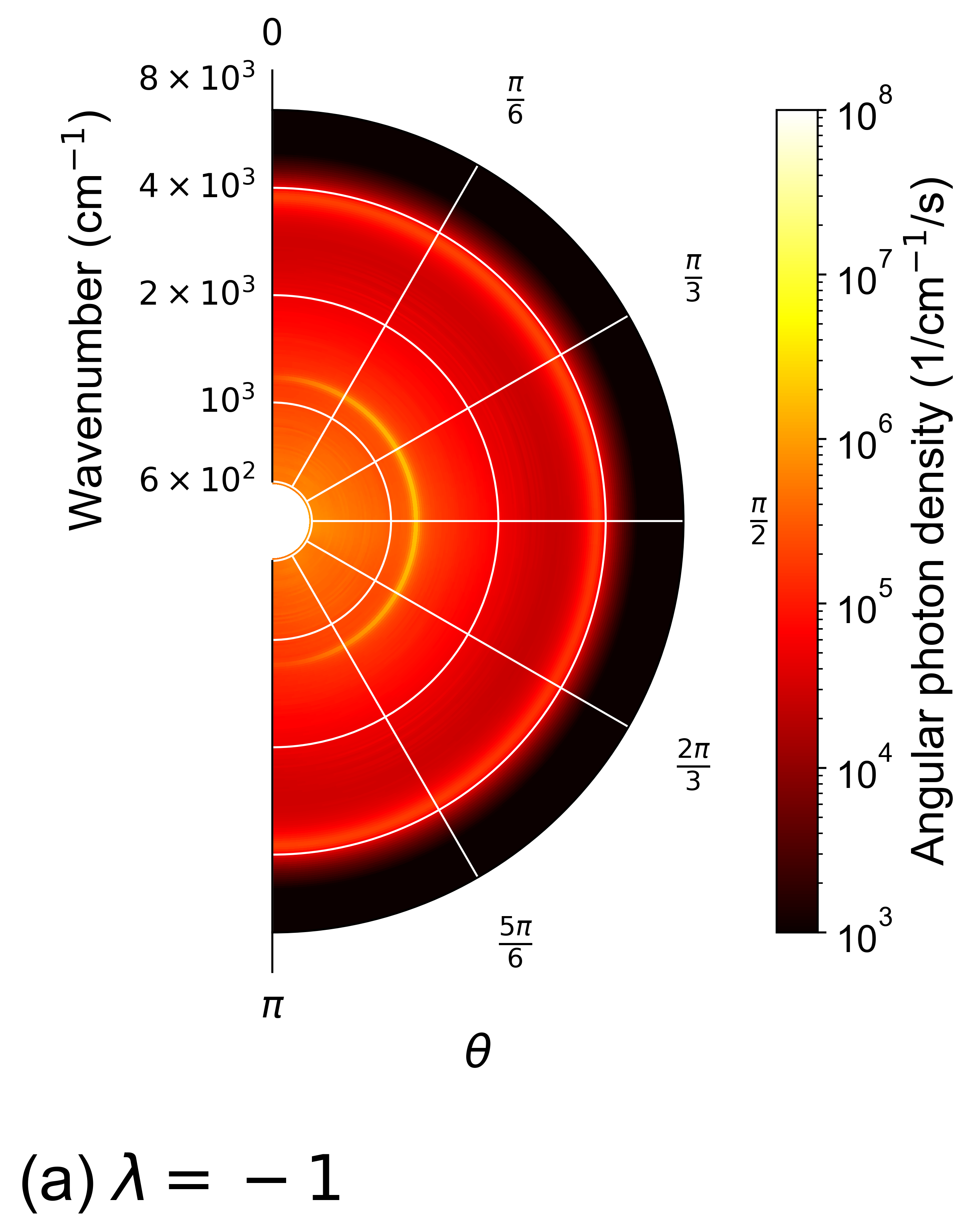}
\includegraphics[width=0.4\linewidth]{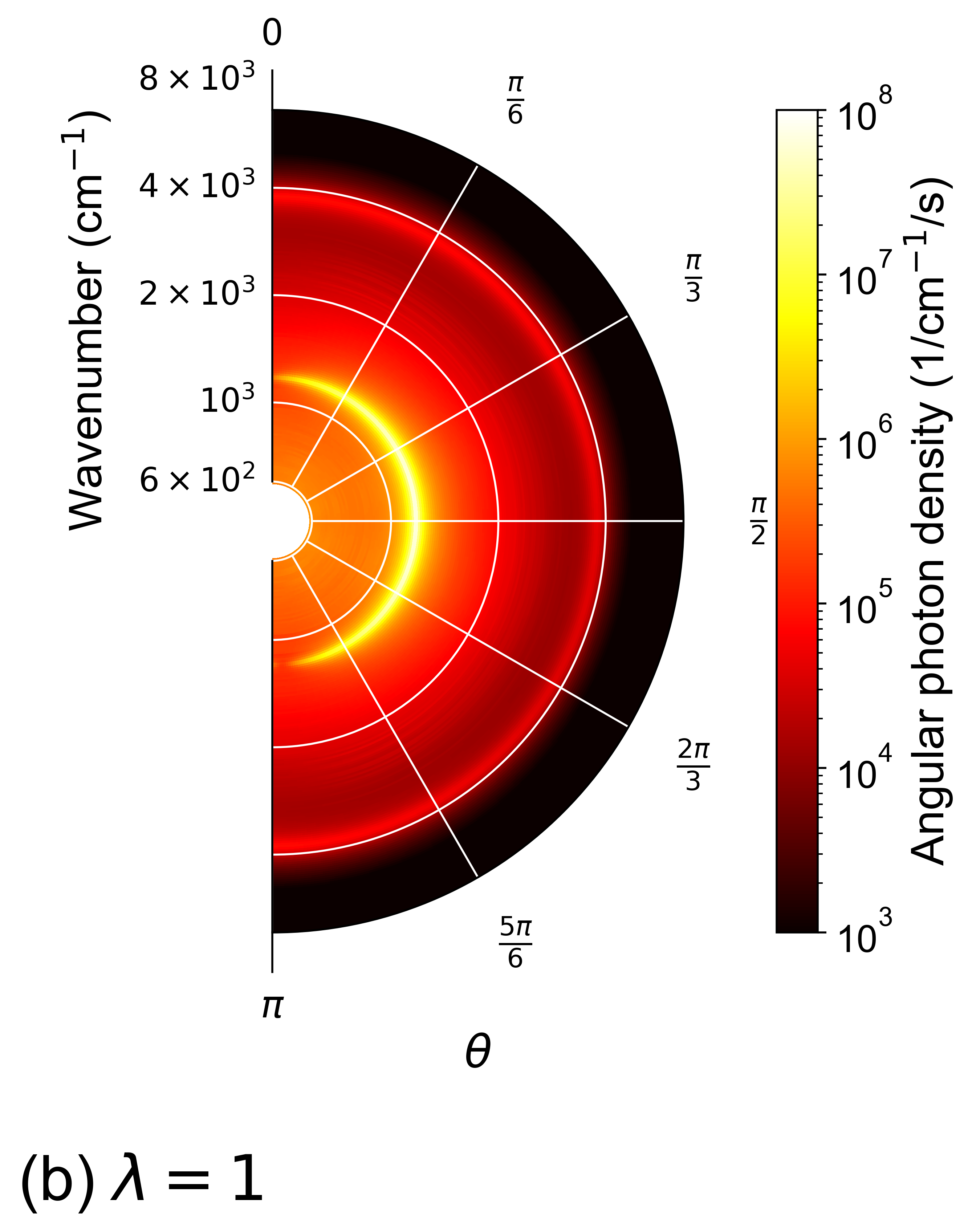}\crcr
\includegraphics[width=0.45\linewidth]{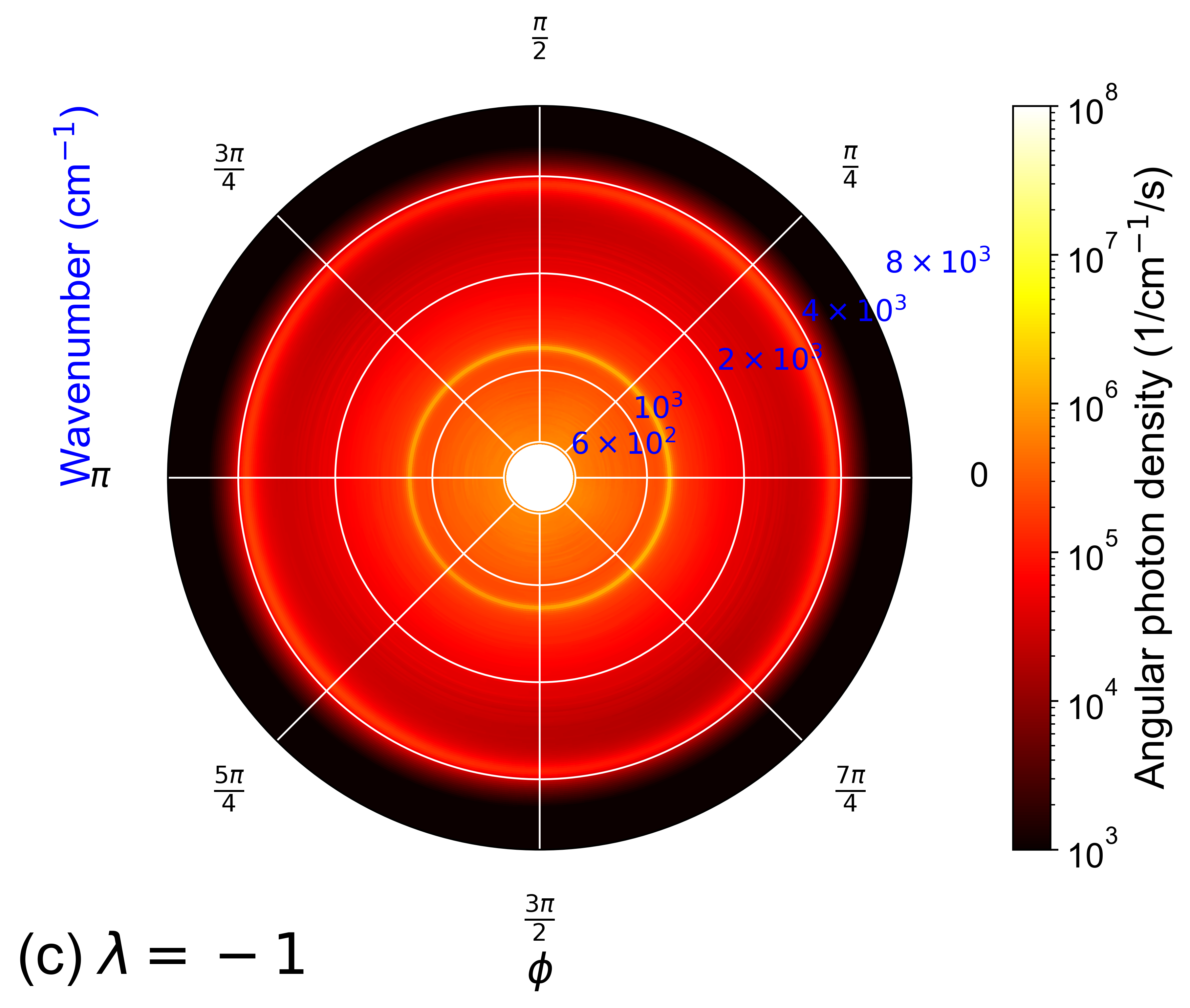}
\includegraphics[width=0.45\linewidth]{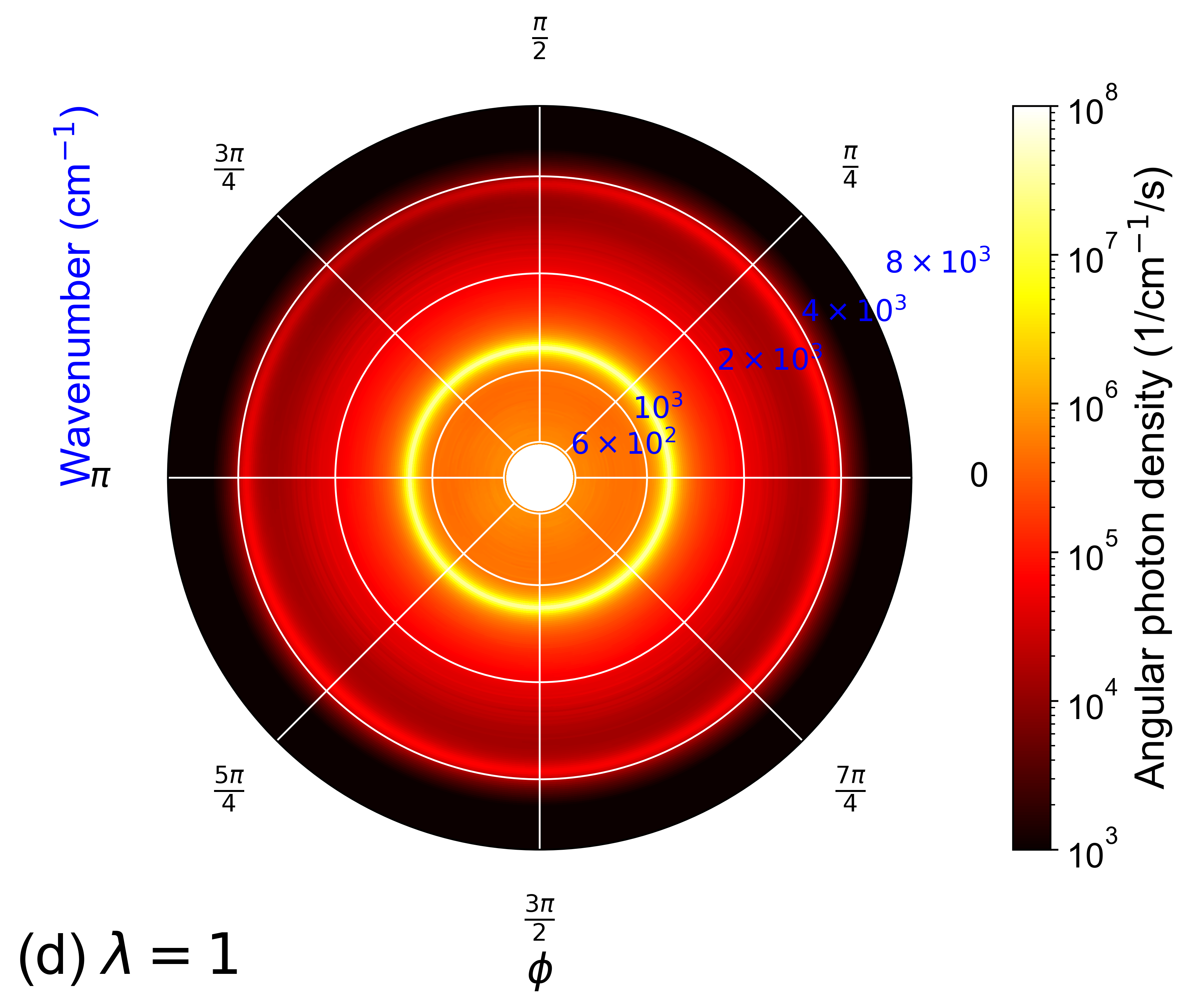}
    \end{tabular}
\end{center}
	\caption{Photon emission rate density as a function of the wavenumber and the polar angle $\theta$ [panels (a) and (b)] for $\phi = \pi/4$, and the azimuthal angle $\phi$ [panels (c) and (d)] for $\theta = \pi/4$. Computations for panels (a),(c) [(b),(d)] were computed with $\lambda = -1$ [$\lambda = 1$].  The structure was illuminated with a Planckian thermal bath at \SI{595}{\K}.}
\label{figAppendix}
\end{figure*}

\section{Angular photon emission rate density of the chiral helix\label{app:helixplots}}
Figure~\ref{figAppendix} shows the photon emission rate density as a function of the wavenumber, the polar angle $\theta$ (panels (a) and (b)) for $\phi = \pi/4$, and the azimuthal angle $\phi$ (panels (c) and (d)) for $\theta = \pi/4$.

\section{Details of the quantum chemistry calculations\label{app:quantumchem}}

All quantum chemistry calculations of the right chiral enantiomer of 1,1'-Bi-2-naphthol (BINOL) molecule were based on the density functional theory (DFT) method and its time-dependent response theory variant (TD-DFT), and performed with a development version of \pkg{TURBOMOLE} electronic structure program. \cite{TURBOMOLE2022, TM_TODAY_TOMORROW} 

The molecular geometry of the R-cis-BINOL was optimized using the gradient minimization technique by iteratively updating positions of the atoms based on classical Newton's equations of motion until forces acting on atoms and change of total electronic energy are below defined threshold. The forces are calculated from the converged electron density in each iteration for a fixed position of atoms. The molecule was placed in a Chloroform solvent that was accounted for implicitly through the conductor-like screening model (COSMO). The relative permittivity of Chloroform was set to 4.711, and a refractive index of 1.4441 was additionally included for optical calculations. The selected exchange-correlation (XC) functional is a combination of hybrid Becke's three parameter exchange and Lee–Yang–Parr (LYP) correlation functional (B3-LYP, B3LYP) \cite{10.1063/1.464913}. The accompanying atomic orbital basis set is triple-\textzeta-quality ``Karls\-ruhe'' def2-TZVP set \cite{weigendBalancedBasisSets2005, eichkornAuxiliaryBasisSets1995, eichkornAuxiliaryBasisSets1997, weigendAccurateCoulombfittingBasis2006} of Gaussian functions with additional functions for polarization describing orbitals of each atom in the BINOL molecule. Additionally, Grimme’s D3 London-dispersion correction \cite{ caldeweyherGenerallyApplicableAtomiccharge2019} was employed together with Becke–Johnson damping (DFT-D3(BJ)) \cite{caldeweyherExtensionD3Dispersion2017} to account for the non-covalent interactions during optimization. Finally, a plethora of additional algorithms, such as resolution-of-identity (RI) \cite{ri}, multipole accelerated resolution-of-identity (marij) \cite{sierkaFastEvaluationCoulomb2003} as well as semi-numerical approach to calculate exchange (senex, esenex) \cite{Holzer2020senex} were employed to speed up calculations without reducing the quality of the obtained results. 

Upon obtaining the structure of the molecule, additional calculations were performed to obtain damped electric-electric, electric-magnetic, and magnetic-magnetic dynamic polarizability tensors stemming from vibrational modes of the molecule, as well as from electronic transitions upon ultraviolet (UV) and visible (Vis.) excitations. The vibrational polarizabilities were calculated for 60.000 equally-spaced frequencies covering the spectral range from 190-10$^7$\,\si{\nano\meter}. The Lorentzian damping was set to \SI{5}{\invcm} for a full-width at half-maximum (FWHM). The dynamic polarizabilities in the UV-Vis. part of the spectrum have been calculated for 411 frequencies in the spectral window from \SI{190}{\nano\meter} to \SI{600}{\nano\meter}. The damping was set to \SI{0.05}{\eV} for FWHM. Finally, to match frequencies from vibrational calculation, we applied the interpolation and extrapolation techniques to obtain dynamic polarizability tensors in the IR part of the electromagnetic spectrum where vibrational modes are active. The dynamic polarizabilities were then used to construct T-matrices separately for vibrational and electronic excitations. The frequency-wise sum of these two kinds of T-matrices was used in Maxwell simulations of thermal emission of the R-BINOL molecule. \cite{FerCor2018}

\bibliography{references}
\end{document}